\pdfoutput=1  
\documentclass[]{article}  
\usepackage{url,float}
\usepackage{graphicx}
\usepackage{amsfonts}
\usepackage{amssymb}
\usepackage{latexsym}

\newcommand{\hide}[1]{}

\newcommand{\ABox}{
\raisebox{3pt}{\framebox[6pt]{\rule{6pt}{0pt}}}
}

\newcommand{\figlab}[1]{\label{fig:#1}}

\newcommand{\eqref}[1]{\ref{eq:#1}}
\newcommand{\figref}[1]{\ref{fig:#1}}

{\makeatletter
 \gdef\xxxmark{%
   \expandafter\ifx\csname @mpargs\endcsname\relax 
     \expandafter\ifx\csname @captype\endcsname\relax 
       \marginpar{xxx}
     \else
       xxx 
     \fi
   \else
     xxx 
   \fi}
 \gdef\xxx{\@ifnextchar[\xxx@lab\xxx@nolab}
 \long\gdef\xxx@lab[#1]#2{{\bf [\xxxmark #2 ---{\sc #1}]}}
 \long\gdef\xxx@nolab#1{{\bf [\xxxmark #1]}}
 \gdef\turnoffxxx{\long\gdef\xxx@lab[##1]##2{}\long\gdef\xxx@nolab##1{}}%
}

\def\P{{\mathcal P}}

\title{Unfolding Orthogonal Terrains}

\author{%
Joseph O'Rourke%
    \thanks{Dept. Comput. Sci., Smith College, Northampton, MA
      01063, USA.
      \protect\url{orourke@cs.smith.edu}.}
}

\begin{document}
\maketitle

\begin{abstract}
It is shown that every orthogonal terrain, i.e., an orthogonal (right-angled)
polyhedron based on a rectangle that meets every vertical line in a segment,
has a grid unfolding: its surface may be unfolded to a single non-overlapping piece
by cutting along grid edges defined by coordinate planes through every vertex.
\end{abstract}

\section{Introduction}
This paper is concerned with \emph{unfolding} the surface of a polyhedron
to a single, connected planar piece that avoids overlap.
We will concentrate on 
\emph{orthogonal polyhedra}: those whose faces meet at angles
that are multiples of $90^\circ$, and whose edges are parallel to
Cartesian $xyz$-axes.
Figure~\figref{ounfolding} shows an \emph{edge unfolding}
\begin{figure}[htbp]
\centering
\includegraphics[width=0.6\linewidth]{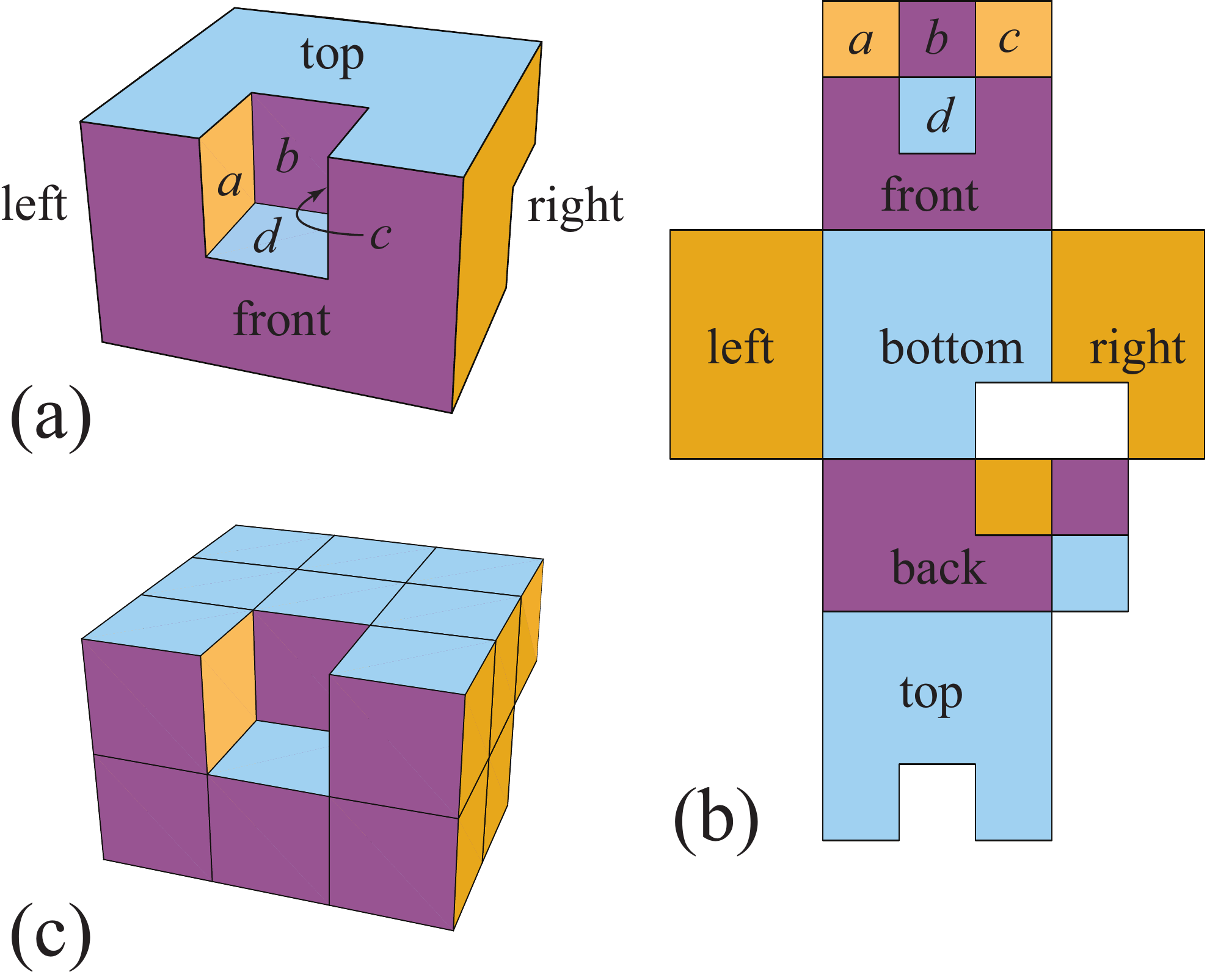}
\caption{(a)~An orthogonal polyhedron.
(b)~An edge unfolding of the polyhedron in~(a).
(c)~Grid edges added to~(a) by intersecting with coordinate planes
through every vertex.}
\figlab{ounfolding}
\end{figure}
of an orthogonal polyhedron, an unfolding produced by cutting
along edges of the polyhedron.
Note that we permit boundary overlap, but no interior points of
the planar piece overlap.
Thus the shape could be cut out of paper and folded up to form the 
surface of the polyhedron.

The study of unfolding orthogonal polyhedra was
initiated in~\cite{bddloorw-uscop-98}, and there are now many
results, which we will not survey 
(see~\cite{o-uop-07} and~\cite{do-gfalop-07}).
It will suffice here to note that an easy example (a small box in
the center of a larger box's top face) demonstrates that
not every orthogonal polyhedron may be edge-unfolded.
Consequently, 
loosenings of the unfolding criteria
have been explored.
A \emph{grid unfolding} adds edges 
(\emph{grid edges}) 
to the surface by intersecting the polyhedron
with planes parallel to Cartesian coordinate planes through every
vertex, as in
Figure~\figref{ounfolding}(c), permitting cutting along these
grid edges.
Even this freedom has not proven sufficient to obtain broadly
applicable algorithms, so
grid refinements have been studied.
A $k_1 {\times} k_2$ \emph{refinement} of a surface~\cite{do-op04-05}
partitions each face into a $k_1 {\times} k_2$ grid of faces
(with the convention that a
$1 {\times} 1$ refinement is an unrefined grid unfolding).
Athough there have now been several grid refinement algorithms
developed that unfold special classes of orthogonal polyhedra
(surveyed in~\cite{o-uop-07}),
it remains unknown whether every orthogonal polyhedron has
a ($1 {\times} 1$) grid unfolding.
This paper shows that a special class of orthogonal polyhedra
does have a grid unfolding.

This class we call \emph{orthogonal terrains}.
Let $P$ be the surface of an orthogonal polyhedron, 
and $\P$ the closed,
solid whose boundary is $P$.
An orthogonal terrain satisfies
two properties:
(1)~there is a distinguished rectangular face of $P$ called the \emph{base} $B$;
and
(2)~every vertical line $L$ (parallel to the $z$-axis) that intersects $\P$
meets it in a single segment, $L \cap \P = s$, with $s$ a finite-length
line segment with one endpoint on $B$: $s \in B$.
$P \setminus B$ is a ``monotone surface,''
and models a terrain of elevations.
In fact, any Digital Elevation Model (DEM), i.e., any rectangular
array of heights, can be viewed as an orthogonal terrain (when closed with
sides and a base).
Figure~\figref{oterrain_10x10} shows an example we will use throughout
the paper (Figure~\figref{ounfolding}(a) is not a terrain because
its base is not a rectangle).
\begin{figure}[htbp]
\centering
\includegraphics[width=0.6\linewidth]{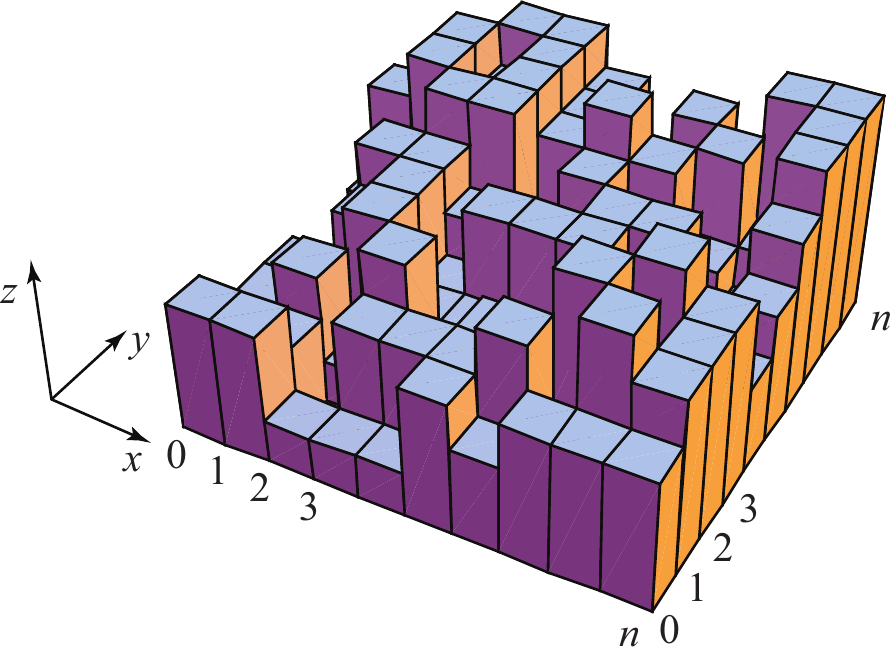}
\caption{A orthogonal terrain with grid edges added,
in this case via a plane at every integer coordinate.
The base $B$ underneath is a $10 {\times} 10$ square.}
\figlab{oterrain_10x10}
\end{figure}

A slightly broader class of shapes,
the ``Manahattan towers,''
were studied in~\cite{dfo-umt-05}.
These differ from terrains only in permitting the base $B$ to be an arbitrary orthogonal
polygon.
This apparently small generalization considerably complicates the situation,
and that paper achieved only a
$5 {\times} 4$ grid unfolding.
Insisting that $B$ be a rectangle permits a completely different, and
relatively simple algorithm
to achieve a $1 {\times} 1$ grid unfolding.

\section{Terrain Unfolding Algorithm}
We now proceed to describe that algorithm, relying on illustrations to
avoid excessive formality.
Unlike most unfolding algorithms, this one can be specified as
a continuous motion that avoids self-intersection throughout 
(as opposed to only guaranteeing nonoverlap at the planar conclusion).
The first two steps are straightforward.
First, the right ($+x$), left ($-x$), and back ($+y$)
vertical faces are unfolded to the $xy$-plane 
while remaining attached to the base $B$.
See Figure~\figref{sides_10x10}.
\begin{figure}[htbp]
\centering
\includegraphics[width=0.95\linewidth]{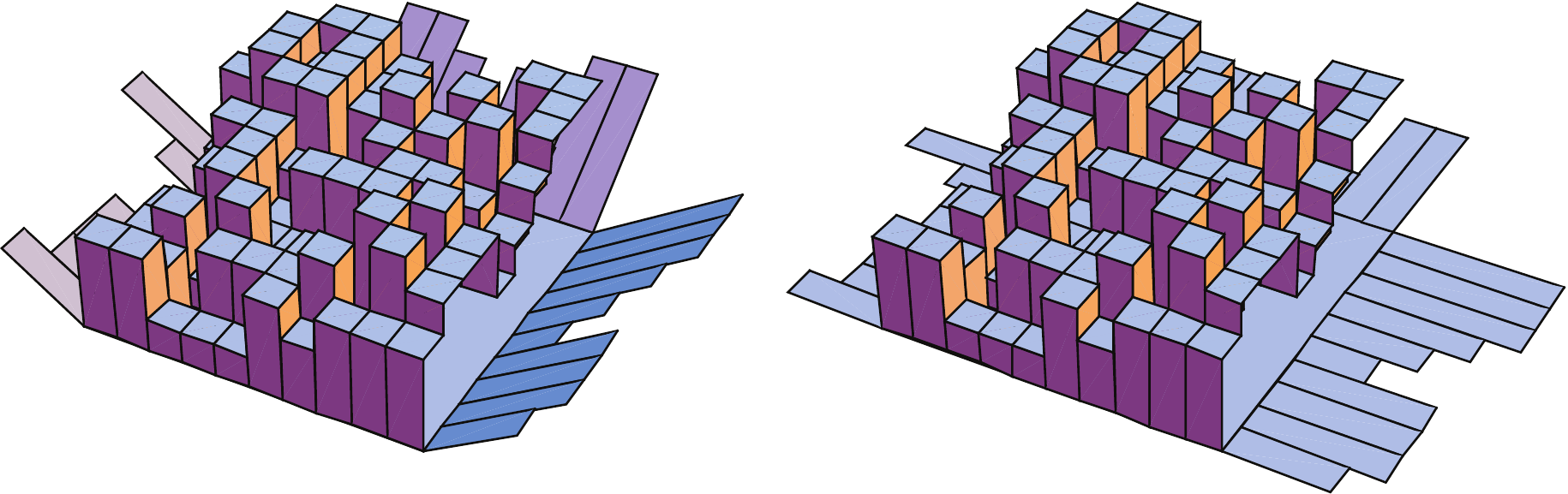}
\caption{Unfolding the right, left, and back sides of $P$.}
\figlab{sides_10x10}
\end{figure}
Second, $B$ and its attachments are rotated around the $x$-axis,
and then the front vertical faces unfolded
horizontally as in
Figure~\figref{front_10x10}.
Here the line of rotation is $x=0 \cap z=h$, where $h$ is the height of
the tallest front face ($h=3$ in the figure; six front faces are tied for
tallest).
\begin{figure}[htbp]
\centering
\includegraphics[width=0.95\linewidth]{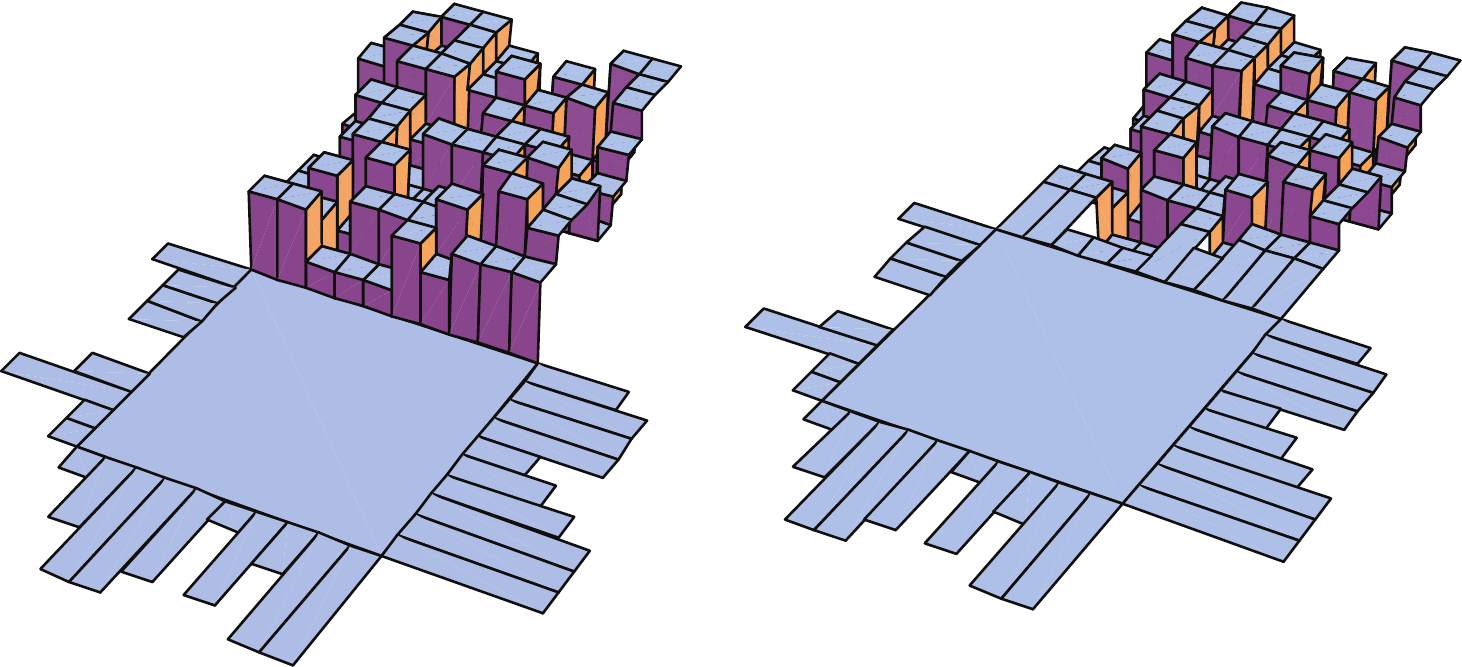}
\caption{Flipping the base $B$ around the line $y{=}z{=}0$,
and then unfolding the front faces of $P$.}
\figlab{front_10x10}
\end{figure}

All this is straightforward.
The third step of the algorithmm is the heart of it.
Define an \emph{$x_i$-strip} as the sequence of faces between
$y=i$ and $y=i+1$ ($i=0,\ldots,n{-}1$) on the ``top'' of $P$:
the horizontal $xy$-faces, and the vertical $yz$ right and left
faces connecting them in a sinuous path.  Each $x_i$-strip will
be unfolded as a unit, into a (long) rectangle stretching in the $x$-direction.
For example, the first $x_0$-strip (covering $y=[0,1]$) in
Figure~\figref{oterrain_10x10} unfolds to
a $16{\times}1$ rectangle: $n=10$ unit square top $xy$-faces, connected by right/left
pairs of $1{\times}2$ and $1{\times}1$ vertical faces.
See Figure~\figref{overhead_10x10}.
\begin{figure}[htbp]
\centering
\includegraphics[width=\linewidth]{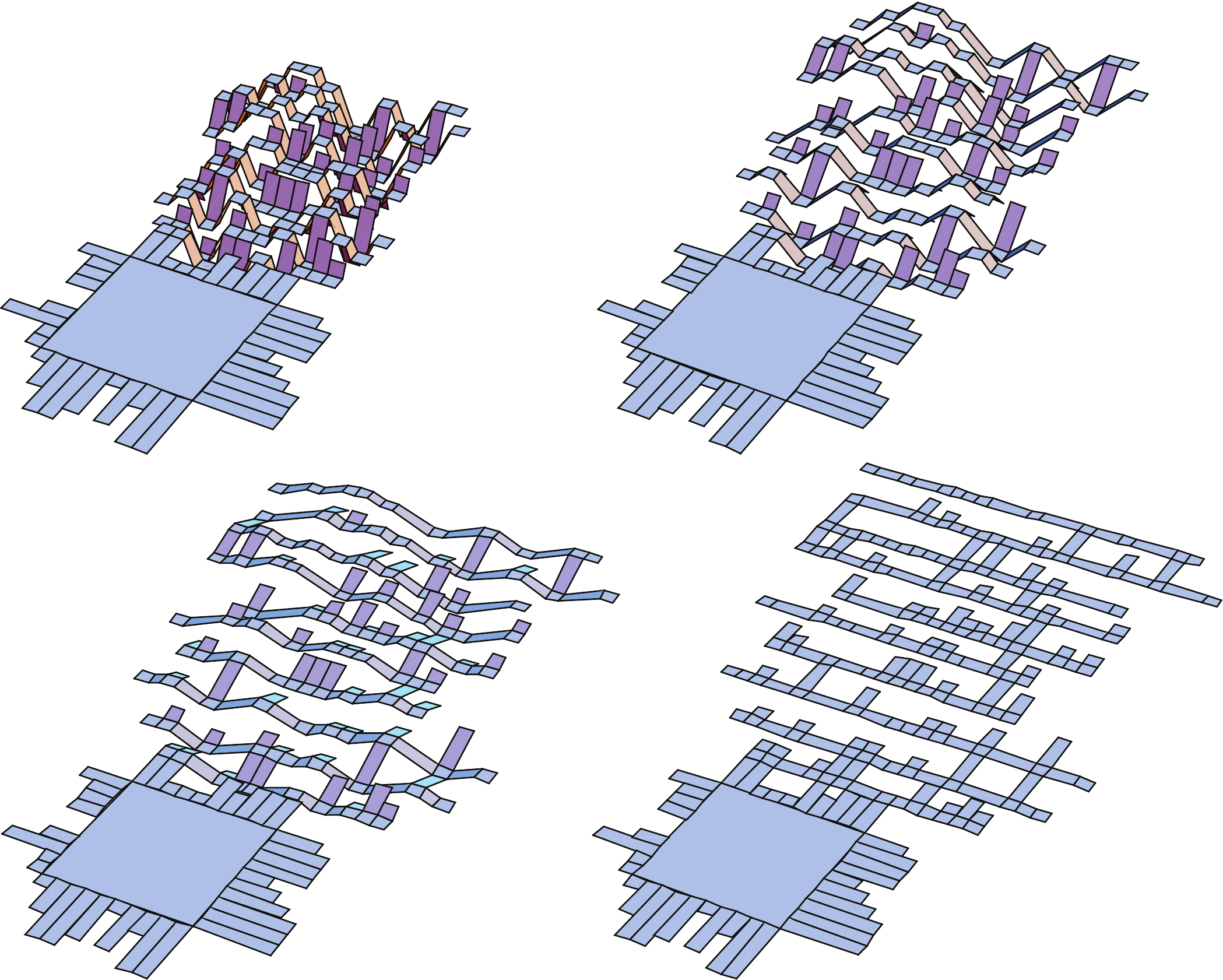}
\caption{Unfolding the top faces of $P$ 
into $x$-strips connected by $y$-bridges.}
\figlab{top_10x10}
\end{figure}

Consider any adjacent $x$-strips, $x_{i-1}$ and $x_i$.
In the original $P$, they are connected by a number of vertical $xz$-faces,
some rising at $y{=}i$ to connect to a higher $y$-adjacent ``tower,'' 
and some descending to connect to a lower $y$-adjacent neighbor.
Define the \emph{bridge} $b_i$ to be the $xz$-rectangle of greatest $z$-height
between the strips,
breaking ties arbitrarily.  Then we lay out the $x_i$-strip 
separated from the $x_{i-1}$ strip 
by the height of $b_i$ in the planar vertical ($y$-) direction, 
and 
aligned horizontally so that $b_i$ connects the two strips.
Note that all the connecting $xz$-rectangles are attached above
the $x_i$ strip.
The continuous unfolding process is depicted in 
Figure~\figref{top_10x10},
and the final unfolding is shown in 
Figure~\figref{overhead_10x10}.
\begin{figure}[htbp]
\centering
\includegraphics[width=0.6\linewidth]{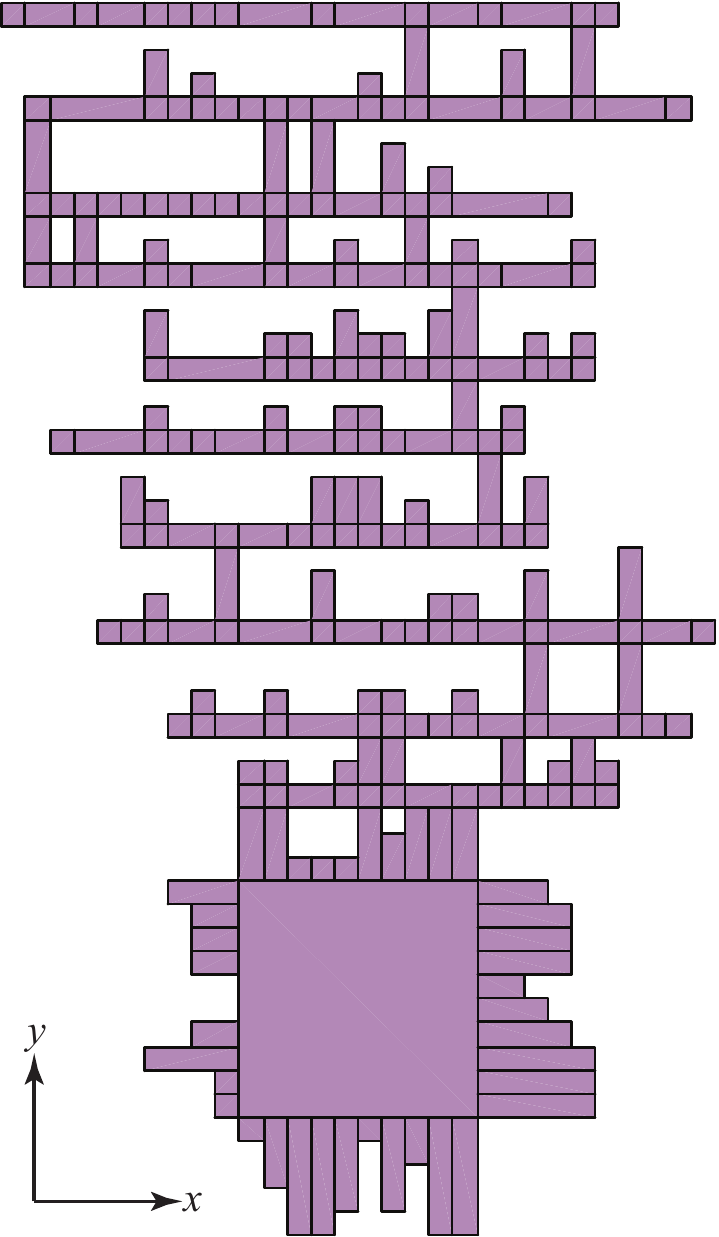}
\caption{The final unfolding of $P$ from Fig.~\protect\figref{oterrain_10x10} in
the $xy$-plane.}
\figlab{overhead_10x10}
\end{figure}
Note that, because of ties, the unfolding is not a simple polygon;
rather, the boundary overlaps at several places.
However, the unfolding is what is known as \emph{weakly simple}, in
that no interior points overlap, as mentioned previously.

\section{Conclusion}

Although our example gridded the polyhedron at every integer lattice point,
it is clear that a coordinate grid plane through every vertex suffices for the
algorithm.

Orthogonal terrains add to the narrow classes of orthogonal polyhedra
that are known to be grid-unfoldable
(orthotubes, 
well-separated orthotrees,
orthogonally convex orthostacks; see~\cite{o-uop-07}),
although
it may be that all orthogonal polyhedra may be grid-unfoldable.
Even extending this new algorithm to terrains defined by slanted axes
(e.g., Figure~\figref{slantedzy})
remains problematical.
\begin{figure}[htbp]
\centering
\includegraphics[width=0.9\linewidth]{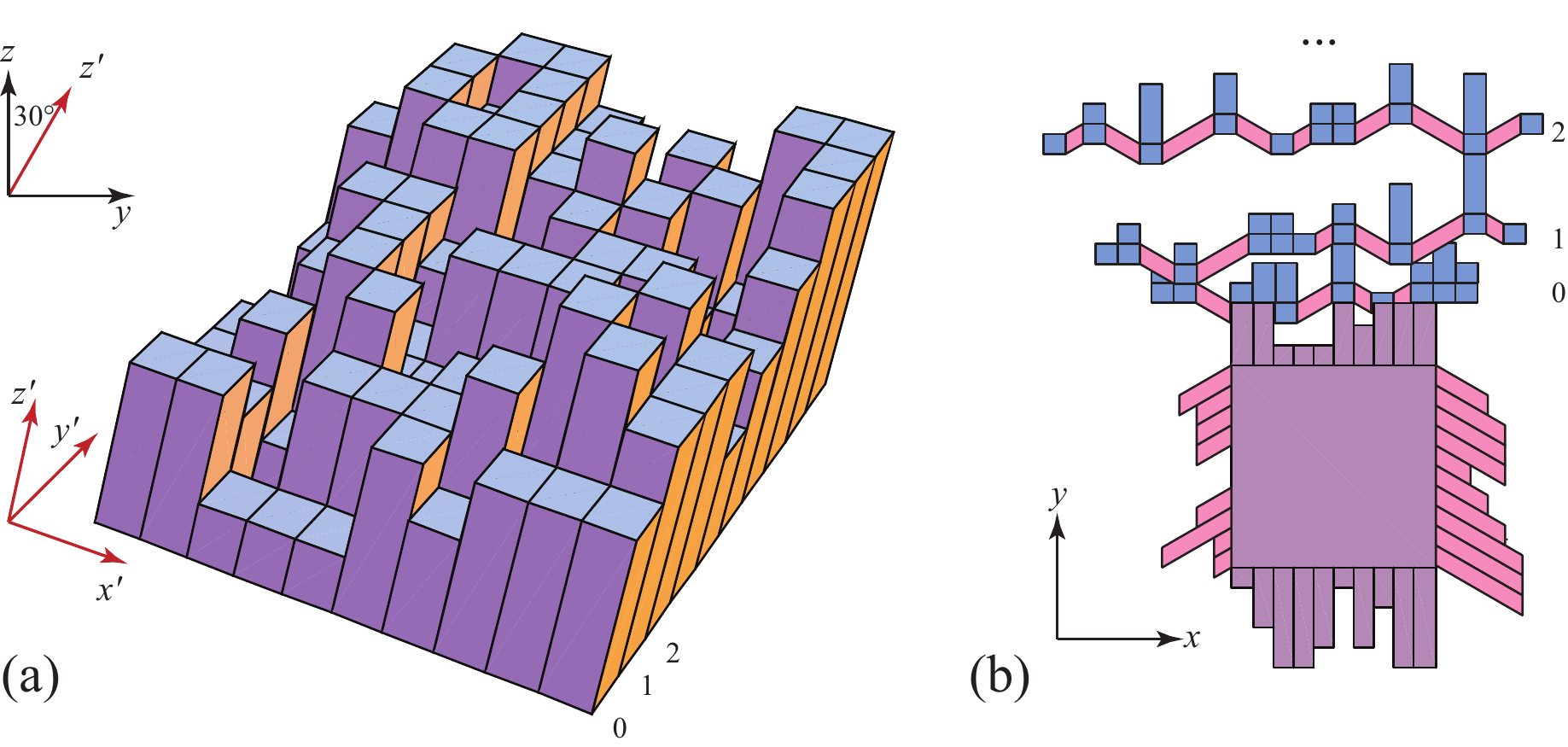}
\caption{(a)~The polyhedron from Fig.~\protect\figref{oterrain_10x10}
with the $z$ axis slanted $30^\circ$ toward the $y$-axis.
(b)~Partial unfolding of first three strips 
$\{x_0,x_1,x_2\}$, showing that the algorithm 
that produced Fig.~\protect\figref{overhead_10x10}
now leads to overlap.
}
\figlab{slantedzy}
\end{figure}

\paragraph{Acknowledgments.}
I am indebted to
Mirela Damian and Robin Flatland, whose work in~\cite{dfo-umt-05}
led to the algorithm in this paper.
I thank Stefan Langerman for the slanted axes question.


\bibliographystyle{alpha}
\bibliography{/home/orourke/bib/geom/geom}
\end{document}